\documentclass[prl,twocolumn,fleqn,showpacs]{revtex4}

\usepackage{graphicx}
\usepackage{bm}
\usepackage{amsmath}

\begin{document}
\newcommand{\Area}{A}
\newcommand{\Av}{{\bm A}}
\newcommand{\av}{{\bm a}}
\newcommand{\Bv}{{\bm B}}
\newcommand{\DOS}{{\nu}}
\newcommand{\dx}{{d^3 x}}
\newcommand{\eF}{{\epsilon_F}}
\newcommand{\etaz}{{\eta^0}}
\newcommand{\etazk}{\eta^0_k}
\newcommand{\etazmk}{\eta^0_{-k}}
\newcommand{\Ev}{{\bm E}}
\newcommand{\js}{{j_{\rm s}}}
\newcommand{\jsc}{{j_{\rm s}^{\rm cr}}}
\newcommand{\jsca}{{j_{\rm s}^{\rm cr (1)}}}
\newcommand{\jscb}{{j_{\rm s}^{\rm cr (2)}}}
\newcommand{\jc}{{j^{\rm cr}}}
\newcommand{\kF}{{k_F}}
\newcommand{\kp}{{k_\perp}}
\newcommand{\Kp}{{K_\perp}}
\newcommand{\kB}{{k_B}}
\newcommand{\kBT}{{k_B T}}
\newcommand{\kv}{{\bm k}}
\newcommand{\Mv}{{\bm M}}
\newcommand{\nv}{{\bm n}}
\newcommand{\pv}{{\bm p}}
\newcommand{\Ne}{{N_{\rm e}}}
\newcommand{\Omegazk}{{\Omega^{(0)}_{\kv}}}
\newcommand{\Omegaok}{{\Omega^{(1)}_{\kv}}}
\newcommand{\omegazk}{{\omega^{(0)}_{\kv}}}
\newcommand{\Omegak}{{\Omega_{\kv}}}
\newcommand{\omegaell}{{\omega_{\ell}}}
\newcommand{\phiz}{{\phi_0}}
\newcommand{\phizc}{\bar{\phi_0}}
\newcommand{\qv}{{\bm q}}
\newcommand{\rv}{{\bm r}}
\newcommand{\rhow}{{\rho_{\rm w}}}
\newcommand{\sigmav}{{\bm \sigma}}
\newcommand{\Sv}{{\bm S}}
\newcommand{\Smh}{{\left(S-\frac{1}{2}\right)}}
\newcommand{\Tv}{{\bm T}}
\newcommand{\torq}{{\tau}}
\newcommand{\Tz}{{\tau}}
\newcommand{\tileta}{{\tilde\eta}}
\newcommand{\tilF}{\tilde{F}}
\newcommand{\tilFc}{\tilde{F}_{\rm c}}
\newcommand{\tilK}{{\tilde{K}}}
\newcommand{\tilKp}{{\tilde{K}_\perp}}
\newcommand{\tiln}{{\tilde{n}}}

\newcommand{\tilTz}{v_{\torq}}
\newcommand{\tilv}{{\tilde{v}}}
\newcommand{\thetaz}{{\theta_0}}
\newcommand{\Vz}{{V_0}}
\newcommand{\xv}{{\bm x}}
\newcommand{\xiz}{\xi_0}
\newcommand{\Vv}{{\bm V}}
\newcommand{\vv}{{\bm v}}
\newcommand{\vc}{v_{\rm c}}
\newcommand{\vck}{v_{\rm c}^{K}}
\newcommand{\vci}{v_{\rm ins}}

Accepted for publication in Physical Review Letters.

\title{Effect of Spin Current on Uniform Ferromagnetism: Domain Nucleation}
\author{Junya Shibata}
\affiliation{Frontier Research System (FRS), The Institute of Physical and 
Chemical Research (RIKEN), 2-1 Hirosawa, Wako, Saitama 351-0198, Japan\\
CREST, JST, 4-1-8 Honcho Kawaguchi, Saitama, Japan}
\author{Gen Tatara}
\affiliation{
PRESTO, JST, 4-1-8 Honcho Kawaguchi, Saitama, Japan \\
Graduate School of Science, Osaka University,
Toyonaka, Osaka 560-0043,
Japan}
\author{Hiroshi Kohno}
\affiliation{
Graduate School of Engineering Science, Osaka University,
Toyonaka, Osaka 560-8531, Japan}
\date{\today}

\begin{abstract}
 Large spin current applied to a uniform ferromagnet 
leads to a spin-wave instability as pointed out recently. 
 In this paper, it is shown that such spin-wave instability is absent 
in a state containing a domain wall, which indicates that nucleation of magnetic domains 
occurs above a certain critical spin current.
 This scenario is supported also by an explicit energy comparison of the 
two states under spin current. 
\end{abstract}
\maketitle

Magnetization dynamics driven by spin torque from a spin polarized current 
(spin current) has been studied extensively after the theoretical prediction 
\cite{{Slonczewski},{Berger96}} 
that a spin current can be used to flip the magnetization in pillar 
(or spin valve) structures 
\cite{{Sun99},{Myers99},{Katine00},{Grollier01},{Tsoi02},{Ralph03}}. 
 As a theoretical framework to describe such current-induced magnetization 
dynamics, Bazaliy, Jones and Zhang (BJZ) \cite{BJZ98} 
derived a modified Landau-Lifshitz-Gilbert (LLG) equation 
for a fully-polarized ferromagnet (half metal) 
with a new term proportional to ${\bm j}\!\cdot\!\nabla{\bm n}$. 
 This new term represents a spin torque that a current of density 
${\bm j}$ exerts on a local magnetization ${\bm n}$.
 Later on, it was generalized to the partially polarized case with an 
interpretation of ${\bm j}$ as the spin current density, ${\bm j}_{\rm s}$ 
\cite{{Macdonald04},Zhang04}. 
 In the Hamiltonian, the effect of spin torque appears in a form 
\begin{equation} 
\label{Spin-Torque}
 H_{\rm ST} = \int d^{3}x \frac{\hbar}{2e}
              {\bm j}_{\rm s}\!\cdot\!\nabla \phi(1-\cos\theta),  
\end{equation} 
where $\theta$ and $\phi$ are polar angles which parameterize ${\bm n}$. 
 As seen from this form, spin current favors a magnetic configuration with 
spatial gradient, or more precisely, with finite Berry-phase curvature. 
 It is thus expected that a large spin current destabilizes a uniform 
ferromagnetic state.
 This is indeed seen from the spin-wave energy \cite{{BJZ98},{Macdonald04},{Zhang04}}, 
\begin{eqnarray}
\label{Omega-uni}
\Omega_{\bm k}^{\rm uni}
=\frac{KS}{\hbar}\sqrt{(k^2\lambda^2+1)(k^2\lambda^2+1+\kappa)} 
+ {\bm k}\cdot{\bm v}_{\rm s},
\end{eqnarray}
where $\lambda = \sqrt{J/K}$ and $\kappa = K_{\perp}/K$, 
with $J$, $K$ and $K_{\perp}$ being exchange constant, 
easy- and hard-axis anisotropy constants, respectively, of localized spins. 
The first term is the well-known spin-wave dispersion with anisotropy gap. 
The effect of spin current appears in the second term as the ``Doppler shift''
\cite{Macdonald04}, where ${\bm v}_{\rm s}(\propto {\bm j}_{\rm s})$ represents 
drift velocity of electron spins. 
 For sufficiently large ${\bm v}_{\rm s}$, $\Omega_{\bm k}^{\rm uni}$ 
becomes negative for a range of ${\bm k}$. 
 This means that there exist states with negative excitation energy, 
indicating the instability of the assumed uniformly-magnetized state 
\cite{{BJZ98},{Macdonald04},{Zhang04}}. 
 The true ground state under spin current, however, remains to be identified. 
 
 In this Letter, we point out that possible ground state is a state 
containing domain walls. 
 Namely, the energy of a domain wall becomes lower than the uniform ferromagnetic state 
 when a spin current exceeds a certain critical value $j_{\rm s}^{\rm cr}$. 
 The behavior of nucleated domains depends much on the magnetic anisotropy 
parameters; 
 Flowing domain wall is stable when the anisotropy is of uniaxial type. 
 For the case of large hard-axis anisotropy, static domain wall is stable 
at the nucleation threshold, but when it starts to flow under larger 
current, even a state with a domain wall becomes unstable. 
 The nature of the resulting state is still unknown. 

Our prediction of domain formation by spin current may be related 
to the very recent experimental observations in metallic and semiconducting 
pillars and films 
\cite{{Ansermet99},{Ji03},{Chen04},{Chen04-2},{Chiba04},{Yang04}}, 
which suggest spatially inhomogeneous magnetization reversal. 
 Domain nucleation is also suggested by recent numerical 
 simulation \cite{Thi04}.

We start by extending the formulation of BJZ \cite{BJZ98} to arbitrary 
degree of polarization, and derive the effective Lagrangian for 
magnetization which is slowly varying in space and time by assumption. 
 With this effective Lagrangian, we calculate spin-wave dispersion around 
a domain wall solution in the presence of spin current. 
 The spin-wave Doppler shift term now has the form, $k(v_{\rm s} - \dot{X})$, 
where $\dot{X}$ is the speed of the domain wall. 
 From this result, we find that the spin-wave instability does not occur 
around a domain wall for any large spin current if the anisotropy is of 
uniaxial type ($K > K_{\perp}$).  
 This suggests that {\it the true ground state is a multi-domain state with 
domain walls}. 
 These domain walls turn out to be flowing with average speed equal to 
drift velocity $v_{\rm s}$ of electron spin determined by the spin current. 
 The stability is then understood from Galilean invariance, since 
a domain wall moving with speed $\dot{X} = v_{\rm s}$ is equivalent 
to a domain wall at rest and in the absence of spin current. 


 The situation is slightly different in the case of large hard-axis 
anisotropy,  $K_{\perp} >\!> K $.
 Domain walls created by the spin current above $j_{\rm s}^{\rm cr}$
are at rest if the current is below depinning threshold, 
$j_{\rm s}^{\rm depin} \propto K_\perp \lambda (>j_{\rm s}^{\rm cr})$.
 At higher current, the walls start to flow, but this triggers another 
spin-wave instability. 
 Thus multi-domain state collapses for $j_{\rm s}>j_{\rm s}^{\rm depin}$ 
into another new ground state which is still unknown \cite{com0}. 
 We defer this new state to future studies. 


We consider a ferromagnet consisting of localized spins and 
conduction electrons.
 The spins are assumed to have an easy $z$-axis and a hard $y$-axis, 
and described by the Lagrangian 
\begin{eqnarray}
\label{LS}
 L_{\rm S} &=& \int \frac{d^3 x}{a^3}\hbar S\dot{\phi}(\cos\theta -1) 
               - H_{\rm S}, \\ 
 H_{\rm S} &=& \int \frac{d^3 x}{a^3}
     \bigg\{\frac{JS^2}{2}(\nabla {\bm n})^2 \nonumber\\
     &&+ \frac{S^{2}}{2}\sin^{2}\theta\left(K+K_{\perp}\sin^{2}\phi\right)
\bigg\}. 
\end{eqnarray} 
 We have adopted a continuum description for localized spins, 
$\Sv({\bm x},t) = S \, {\bm n}({\bm x},t)$, with unit vector 
${\bm n}({\bm x},t) = (\sin\theta\cos\phi, 
\sin \theta\sin \phi, \cos\theta)$ and the magnitude of spin, $S$. 
 The $J$ is the exchange constant, and $a$ the lattice constant. 
 The easy-axis (${K}$) and hard-axis (${K}_\perp$) anisotropy constants 
generally incorporate the effect of demagnetizing field. 
 The exchange interaction between localized spins and conduction electrons 
is given by 
\begin{equation}
\label{Hsd}
\hat{H}_{\rm sd} 
= - \Delta \int \dx \, {\bm n}({\bm x},t)\!\cdot\! (\hat{c}^\dagger({\bm x},t) 
\sigmav \hat{c}({\bm x},t)), 
\end{equation}
where $2\Delta$ and $\hat{c}$ ($\hat{c}^\dagger$) are the energy splitting 
and
annihilation (creation) operator of conduction electrons, respectively,
and $\sigmav$ is a Pauli-matrix vector.
 The free-electron part is given by
$\hat{H}_{\rm el}^{0} 
= \sum_{\kv\sigma} \epsilon_{\kv} 
\hat{c}^\dagger_{\kv\sigma}\hat{c}_{\kv\sigma}$
with $\epsilon_{\kv} = \hbar^2 \kv^2 /2m$. 
 We perform a local gauge transformation 
\cite{{TK04},{TF94}} in electron spin space so that the quantization axis
is parallel to ${\bm S}({\bm x},t)$ at each point of space and time; 
$\hat{c}({\bm x},t) = U({\bm x},t)\hat{a}({\bm x},t)$, 
where $\hat{a}({\bm x},t)$ is the two-component electron operator 
in the rotated frame, and 
$U({\bm x},t) \equiv {\bm m}({\bm x},t)\!\cdot\!{\bm \sigma}$ 
is an SU(2) matrix with 
${\bm m}({\bm x},t) 
 = (\sin(\theta/2)\cos\phi,\sin(\theta/2)\sin\phi,\cos(\theta/2))$. 
 The electron part of the Hamiltonian is now given by 
$\hat{H}_{\rm el}
 = \sum_{\kv\sigma} \epsilon_{\kv\sigma} a^\dagger_{\kv\sigma}a_{\kv\sigma}$ 
with 
$\epsilon_{\kv\sigma} = \hbar^2 \kv^2 /2m -\sigma\Delta$, 
and the interaction Hamiltonian $\hat{H}_{\rm int}$ by 
\begin{eqnarray}
\label{Hint}
\hat{H}_{\rm int} &=& \frac{\hbar}{V}\sum_{\kv\qv\alpha\beta}
\left\{
A_{0}^{\nu}(\qv,t)+\frac{\hbar(2k_{j}+q_{j})}{2m}A_{j}^{\nu}(\qv,t)\right\} 
\nonumber\\
&\times&\hat{a}^{\dagger}_{\kv+\qv\alpha}\sigma^{\nu}_{\alpha\beta}\hat{a}_{\kv\beta} 
\nonumber\\ 
&+&\frac{\hbar^{2}}{2mV^{2}}\sum_{\kv\qv\pv\alpha}A^{\mu}_{j}(\pv,t)A^{\mu}_{j}(\qv-\pv,t)
\hat{a}^{\dagger}_{\kv+\qv\alpha}\hat{a}_{\kv\alpha}. 
\end{eqnarray}
 Here, 
${\bm A}_{\mu}(\qv,t)=\int d^3 x e^{-i\qv\cdot{\bm x}}
  {\bm A}_{\mu}({\bm x},t)$ 
with ${\bm A}_{\mu} ({\bm x},t) = {\bm m}\times \partial_{\mu}{\bm m}$
represents SU(2) gauge field with space ($\mu = x,y,z$) and 
time ($\mu = 0$) components.

 For slowly varying magnetic configurations, the effective Lagrangian can 
be derived by a perturbative expansion with respect to ${\bm A}_{\mu}$.  
 This is the gradient expansion. 
 In conformity with $L_S$, we retain terms up to second order 
in spatial gradient or in ${\bm A}_{i}$ ($i=x,y,z$), 
and first order in time derivative or in ${\bm A}_{0}$. 
 The spin-torque term, 
$\sum_{\bm x} \frac{\hbar}{e}{\bm j}_{\rm s} \!\cdot\! {\bm A^z}$, 
arises from the first-order contribution $\langle \hat{H}_{\rm int} \rangle$.
 Here ${\bm j}_{\rm s} = \frac{e}{V}\sum_{\kv\sigma}\sigma 
\frac{\hbar\kv}{m}f_{\kv\sigma}$ 
is the spin current density in the rotated frame, 
with a distribution function $f_{\kv\sigma} = 
\langle\hat{a}^{\dagger}_{\kv\sigma}(t)\hat{a}_{\kv\sigma}(t)\rangle$ 
specifying the state carrying current. 
 Other terms just give renormalizations of $J$ and $S$ \cite{TF94}. 
 The effective Lagrangian is thus given by 
\begin{eqnarray}
\label{effL}
 L_{\rm eff} = L_{\rm S}-H_{\rm ST}. 
\end{eqnarray}
The second term, given by Eq. (\ref{Spin-Torque}), with $j_{\rm s}$ 
is identical to that derived by BJZ \cite{BJZ98} for the half-metallic case. 
It is seen that spin current favors a finite Berry phase curvature along 
the current. 
 The modified LLG equation can be obtained by taking variations of 
$L_{\rm eff}$ with respect to $\theta$ and $\phi$. 
 It has the same form as in BJZ, and is applicable to general 
slowly-varying spin texture, and to the case of arbitrary degree 
of polarization. 

 Let us now study the spin-wave excitation around a domain-wall solution 
under spin current by using the method of collective coordinate 
\cite{Rajaraman}. 
 The wall is assumed to be planar and moving in the direction perpendicular 
to the plane (chosen as $x$-direction).
 It is convenient to use a complex field 
$\xi(x,t)=e^{i\phi(x,t)}\tan \frac{\theta(x,t)}{2}$ 
to represent magnetization. 
 We decompose it into the domain-wall configuration and spin waves 
around it: 
\begin{eqnarray}
\label{CDF}
\xi(x,t)&=&
\xi_{\rm dw}(x,t)\exp\bigg[2\cosh\{(x-X)/\lambda\}\nonumber\\
&&\times\frac{1}{\sqrt{L}}\sum_{k}\varphi_{k}(x-X)\eta_{k}(t)\bigg], 
\end{eqnarray}
where $\xi_{\rm dw}=\exp[-(x-X(t))/\lambda + i \phi_{0}(t)]$ 
represents a domain-wall configuration. 
 The wall position $X(t)$ and the angle $\phi_{0}(t)$ between wall spins 
and the easy plane are proper collective coordinates 
\cite{{TK04},{BL96},{TT96},{ST00},{ST01}}. 
 The spin-wave part is expanded with modes 
$\varphi_{k}(x) = \frac{1}{\sqrt{k^{2}\lambda^{2}+1}}
(-ik\lambda + \tanh \frac{x}{\lambda})e^{ikx}$ \cite{{BL96},{ST01}}, whose amplitude 
$\eta_{k}$ precisely corresponds to the Holstein-Primakoff boson. 
($L$ is the length of the system in the $x$-direction.) 
 Substituting (\ref{CDF}) into (\ref{effL}), 
we obtain the spin-wave Lagrangian $L_{\rm sw}$, 
up to the quadratic order in $\eta$, as 
\begin{eqnarray}
&&L_{\rm sw} = \frac{\hbar N{S}}{2\lambda}
\sum_{k}\bigg\{i(\eta^{*}_{k}\dot{\eta}_{k}-\dot{\eta}^{*}_{k}\eta)\nonumber\\ 
&& \ \ \ \ \ - 2\left\{\Omega^{(0)}_{k}-k(\dot{X}-v_{\rm s})\right\}
        \eta^{*}_{k}\eta_{k}  \nonumber \\
&& \ \ \ \ \ + \frac{v_{\rm c}}{\lambda}(1-4\sin^{2}\phi_{0})
        (\eta_{-k}\eta_{k}+\eta^{*}_{-k}\eta^{*}_{k})
\bigg\}, 
\label{spinwave}
\end{eqnarray}
where $N=2\lambda A/a^{3}$ is the number of spins in the wall
($A$ is the cross-sectional area of the wall), 
$\Omega^{(0)}_{k} = 
\frac{KS}{\hbar}(k^{2}\lambda^{2}+1+\kappa/2)$ and 
$v_{\rm c}=\frac{\lambda K_{\perp}S}{2\hbar}$. 
 Details of the calculation will be presented in a separate paper. 
 From Eq. (\ref{spinwave}), we find the spin-wave dispersion around a 
current-driven domain wall: 
\begin{eqnarray}
\label{dispersion-dw}
\Omega^{\rm dw}_{k} &=& \frac{KS}{\hbar}
\bigg\{(k^2\lambda^{2}+1)(k^2\lambda^{2}+1+\kappa)\nonumber\\
&+&2\kappa^{2}\sin^{2}\phi_{0}\cos 2\phi_{0}\bigg\}^{1/2}
+k\left(v_{\rm s}-\dot{X}\right).
\end{eqnarray} 
 The most important difference from the case of uniform ferromagnet  is that 
the drift velocity $v_{\rm s} = \frac{a^3}{2eS} j_{\rm s}$ of spin current 
appears as a relative velocity, $v_{\rm s}-\dot{X}$, with respect to the 
moving wall. 

 Equation (\ref{dispersion-dw}) shows that the spin-wave energy depends 
strongly on the wall dynamics (through $\dot{X}$ and $\phi_{0}$).
 From the equations of motion for the domain wall \cite{TK04}, 
$\lambda\dot{\phi}_{0} + \alpha \dot{X} = 0$ and 
$\dot{X}-\alpha \lambda  \dot{\phi_{0}}
= v_{\rm c}\sin 2\phi_{0} + v_{\rm s}$, 
where $\alpha \ll 1 $ is a damping parameter, we have 
$ ( 1 + \alpha^2 )\dot X = v_{\rm c}\sin 2\phi_{0} + v_{\rm s}$. 
 Neglecting terms of $\sim O(\alpha^2)$, we have, 
\begin{eqnarray}
\label{dispersion-dw2}
\Omega^{\rm dw}_{k}& =& \frac{KS}{\hbar}\bigg[
\bigg\{(k^2\lambda^{2}+1)(k^2\lambda^{2}+1+\kappa)\nonumber\\
&+&2\kappa^{2}\sin^{2}\phi_{0}\cos 2\phi_{0}\bigg\}^{1/2}
-\frac{k\lambda}{2}\kappa\sin 2\phi_{0}\bigg],
\label{swenergy}
\end{eqnarray}
which depends on $j_{\rm s}$ implicitly through $\phi_{0}$. 

 For $K > K_{\perp}$, it is easy to see that $\Omega^{\rm dw}_{k}$ 
remains positive for all values of $k$ and $j_{\rm s}$ \cite{com2}.
 Thus a domain-wall state is stable irrespective of the magnitude of 
the spin current in this case.
 This indicates that a uniform ferromagnetism collapses into a 
multi-domain configuration for $j_{\rm s} > j_{\rm s}^{\rm cr}$, 
where 
$j_{\rm s}^{\rm cr}=\frac{2eS^2}{\hbar a^{3}}K\lambda(1+\sqrt{1+\kappa})$ 
is the critical spin current density for the instability of a uniform 
ferromagnet \cite{{BJZ98},{Macdonald04},{Zhang04}}. 
 Since the domain wall flows for 
$j_{\rm s} > j_{\rm s}^{\rm depin} \equiv \frac{eS^2}{\hbar a^{3}} 
   K \lambda \kappa $ \cite{TK04}, 
which satisfies  $j_s^{\rm cr} >j_{\rm s}^{\rm depin}$, 
it starts to flow as soon as it is created (Fig.1(a)).
 The stability of moving domain wall state is natural from Galilean 
invariance since domain wall with velocity of $v_{\rm s}$ 
is equivalent to the static domain wall without spin current.

In the opposite case, $K_{\perp}> K$, the spin wave 
shows no anomaly as long as $ j_{\rm s} < j_{\rm s}^{\rm depin} $,
where the wall remains static with constant $\phi_0$ 
($ =-\frac{1}{2} {\rm sin}^{-1} (v_{\rm s}/v_{\rm c})$). 
 Hence, for $j^{\rm cr}_{\rm s} < j_{\rm s} < j^{\rm depin}_{\rm s}$ 
($j^{\rm cr}_{\rm s} < j^{\rm depin}_{\rm s}$ is realized when 
$K_{\perp} > 8 K$), the static multi-domain state is realized. 
 In contrast, as soon as the wall starts to move 
 under larger spin current ($j_{\rm s} > j_{\rm s}^{\rm depin}$), 
the spin wave becomes unstable (Eq.(\ref{swenergy})). 
 Thus another ground state would appear.  
 The wavelength of the unstable mode around the uniform magnetization, 
$k=-(K(K+K_\perp)/J^2)^{1/4}$\cite{Zhang04}, suggests that this unknown 
ground state has a spatial structure with short length scale of 
$\sim (K K_\perp/J^2)^{-1/4} = (K/K_{\perp})^{1/4}\lambda < \lambda $. 

We have thus shown that the spin-wave instability under spin current 
is avoided by the existence of domain wall. 
 This may indicate that domain nucleation occurs under spin current. 
 We cannot, however, answer here how many domains are created. 
 The above argument holds for each segment larger than 
the wall thickness, $\lambda$, and 
thus the domain wall can be nucleated with spacing of $\sim\lambda$.
 For a correct estimate of the spacing, however, we need to consider 
the dipolar interaction among domains. 
 We will not pursue this point in this letter.

We here present a supporting argument for the above scenario of domain 
formation by evaluating explicitly the total energy of a domain wall 
in the presence of spin current. 
 We consider only the case $K \ll K_\perp$ and
$j_{\rm s}^{\rm cr} < j_{\rm s} \ll j_{\rm s}^{\rm depin}$, where the
wall can remain static and thus the energy comparison has physical
meaning. 
 A static domain wall configuration is given by 
$\theta(x,t)=\theta_{\rm s}(x)=2\tan^{-1}\exp(-x/\lambda')$, 
$\phi(x,t)=\phi_{\rm s}(x)$, where $\lambda'=\lambda/\sqrt{1+\kappa\sin^{2}\phi_{0}}$ 
includes the effect of contraction of domain wall \cite{BKL97} 
($\phi_{0} \equiv -\frac{1}{2}\int dx \phi_{s}(\partial_{x}\theta_{\rm s})\sin\theta_{\rm s}$). 
Note that $\phi_{\rm s}(x)$ is not spatially constant when there is a spin current. 
 For large hard-axis anisotropy, $K_{\perp} \gg \frac{\hbar 
a^{3}}{eS^{2}\lambda}j_{\rm s}$, 
substitution of the static solution $\theta_{\rm s}$ and $\phi_{\rm s}$ into the 
modified LLG equation 
leads to 
$ \phi_{0} \simeq - (\hbar a^{3}/2eS^{2}\lambda K_{\perp}) j_{\rm s} 
  \ll 1$. 
The total energy, $H_{\rm S}+H_{\rm ST}$, of this static domain wall is then given by \cite{partial}\begin{eqnarray}
 E_{\rm dw}&=&H_{\rm S}|_{\theta_{\rm s},\phi_{\rm s}}
 -\frac{\hbar}{2e}j_{\rm s}
 \int d^{3}x \phi_{\rm s}\partial_{x}\theta_{\rm s}\sin\theta_{\rm s}\nonumber\\
  &\simeq& NKS^{2}\left[1-\frac{1}{2} 
  \left(\frac{\hbar a^{3}}{2eS^{2}\sqrt{KK_{\perp}}\lambda} j_{\rm s} 
       \right)^{2}\right]. 
\end{eqnarray}
 Since the energy $E_{\rm uni}$ of the uniform state is zero independently 
of $j_{\rm s}$, we find that $E_{\rm dw}< E_{\rm uni}$ when 
$j_{\rm s}>\frac{2eS^2}{\hbar a^{3}}\sqrt{2KK_{\perp}}\lambda$ (Fig.1(b)). 
 This critical value is of the same order of magnitude as $j_{\rm s}^{\rm cr}$ 
 for $K_{\perp}\gg K$. 
 Thus, the uniformly magnetized state is a `false vacuum', 
which collapses into a multi-domain state via first order phase transition 
under spin current above a critical value. 

The domain formation may be understood from the spin wave 
around uniform ferromagentism.
 In fact, the spin wave energy,  
$\Omega^{\rm uni}_{\bm k}$, given by Eq. (\ref{Omega-uni}),  
indicates that the typical wavelength of the spin wave excitation 
(determined by the minimum of the spin wave energy) is given by 
$k^{-1}\sim -\lambda^{2}(\sqrt{1+\kappa}KS)/(\hbar v_{\rm s})$ for small $j_{\rm s}$.
 As the current increases, the wavelength becomes shorter and the 
critical current is given by the condition it becomes
comparable to the wall thickness.

The above estimate of the critical current density obtained for a
 planar wall would be applicable also to the
case of nucleation in films such as in Ref.\cite{Chen04-2}, at least 
for an estimate of order of magnitude.

Let us estimate the critical current numerically. 
 The magnitude of the current needed for nucleation in a wire
is estimated as $j_{\rm s} \sim 3~\times~10^{13}~{\rm A}/{\rm m}^{2}$ 
by using the materials parameters of Co:
$a= 2.3$~\AA,
$S \sim 1$, $J= 3~\times~10^{-40}~{\rm J}{\rm m}^{2}$,
$K= 6.2~\times~10^{-24}~{\rm J}$,
($\lambda=\sqrt{J/K}=76~$\AA), under the assumption that 
$K_{\perp}$ is mostly due to the demagnetizing field,
$K_{\perp} = M^{2}_{\rm s}a^{3}/\mu_{0} = 3~\times~10^{-23}~{\rm J}$. 
 Recent experiment on a Co layer with a current through a point contact with 
domain
suggests a domain wall formation for a maximum current density close to the 
contact region of
$j\sim 5 \times 10^{12-13}$ A/${\rm m}^{2}$ 
\cite{{Ji03},{Chen04},{Chen04-2}}. 
 This value is in rough agreement with our estimate.
 Other recent experiments on metallic and semiconducting pillar structures 
indicating inhomogeneous magnetization reversal may also 
be due to domain (or domain-like) structure 
created by spin current \cite{{Ansermet99},{Chiba04},{Yang04}}.

The authors would like to thank T. Yang, T. Kimura, H. Ohno, 
T. Ono and Y. Otani 
for showing experimental data and valuable discussions. 
This work is partially supported by Grant-in-Aid of Monka-sho, Japan. 
G.T. thanks Mitsubishi foundation for financial support.



\begin{figure}[bp]
\includegraphics[scale=0.4]{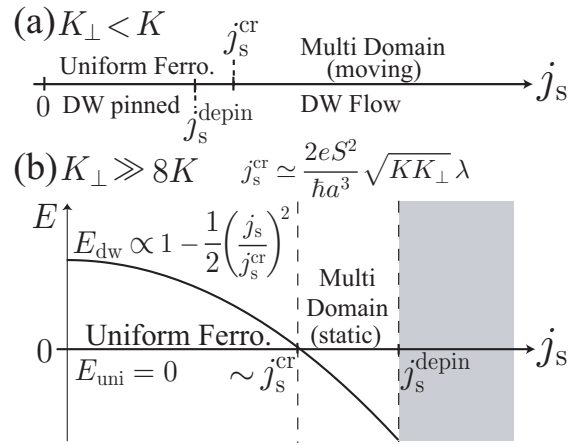}
\caption{Schematic phase diagram under spin current $j_s$ in the absence 
of pinning potential.  
(a) $K_{\perp} < K$. Above $j_{\rm s}^{\rm cr}$, uniform ferromagnetism 
collapses into multi-domain structure in which domain walls are flowing 
due to spin current. 
 The threshold, $j_{\rm s}^{\rm depin}$, for "depinning" from $K_\perp$ 
 is below $j_{\rm s}^{\rm cr}$. 
(b) $K_{\perp} \gg 8K$. 
 Energy of the single-wall state ($E_{\rm dw}$) is compared with 
 that of the uniformly magnetized state, $E_{\rm uni} = 0$. 
 Multi-domain state here remains at rest. In the gray region, 
($j_{\rm s}>j_{\rm s}^{\rm depin}$), the domain wall starts to flow 
but is unstable, suggesting a new ground state.}
\label{fig}
\end{figure}


\begin{thebibliography}{99}

\bibitem{Slonczewski}
J. C. Slonczewski, J. Magn. Magn. Mater. {\bf 159}, L1 (1996). 

\bibitem{Berger96}
L. Berger, Phys. Rev. B {\bf 54}, 9353 (1996). 


\bibitem{Sun99}
J. Z. Sun, J. Magn. Magn. Mater. {\bf 202}, 157 (1999).

\bibitem{Myers99}
E. B. Myers, D. C. Ralph, J. A. Katine, R. N. Louie, and Buhrman, 
Science {\bf 285}, 867 (1999). 

\bibitem{Katine00}
J. A. Katine, F. J. Albert, R. A. Buhrman, E. B. Myers, and D. C. Ralph, 
Phys. Rev. Lett. {\bf 84}, 3149 (2000). 

\bibitem{Grollier01}
J. Grollier, V. Cros, A. Hamzic, J. M. George, H. Jaffr\`es, A. Fert, 
G. Faini, J. Ben Youssef, and H. Legall, 
Appl. Phys. Lett. {\bf 78}, 3663 (2001). 

\bibitem{Tsoi02}
M. Tsoi, V. Tsoi, J. Bass, A. G. M. Jansen, and P. Wyder, 
Phys. Rev. Lett. {\bf 89}, 246803 (2002). 

\bibitem{Ralph03}
S. I. Kiselev, J. C. Sankey, I. N. Krivorotov, N. C. Emley, R. J. 
Schoelkopf, R. A. Buhrman, and D. C. Ralph, 
Nature (London) {\bf 425}, 380 (2003). 


\bibitem{BJZ98}
Ya. B. Bazaliy, B. A. Jones, and Shou-Cheng Zhang, 
Phys. Rev. B {\bf 57}, R3213 (1998). 


\bibitem{Macdonald04}
J. Fern\'andez-Rossier, M. Braun, A.S. N\'u\~nez and A. H. MacDonald, 
Phys. Rev. B {\bf 69}, 174412 (2004). 

\bibitem{Zhang04}
Z. Li and S. Zhang, Phys. Rev. Lett. {\bf 92}, 207203 (2004). 


\bibitem{Ansermet99}
J.-E. Wegrowe, D. Kelly, Y. Jaccard, 
Ph. Guittienne, and J.-Ph. Ansermet, 
Europhys. Lett. {\bf 45}, 626 (1999). 






\bibitem{Ji03}
Y. Ji, C. L. Chien, and M. D. Stiles, 
Phys. Rev. Lett. {\bf 90}, 106601 (2003). 

\bibitem{Chen04}
T. Y. Chen, Y. Ji, and C. L. Chien, Appl. Phys. Lett. {\bf 84}, 380 (2004). 

\bibitem{Chen04-2}
T. Y. Chen, Y. Ji, C. L. Chien, and M. D. Stiles, 
Phys. Rev. Lett. {\bf 93}, 026601 (2004). 

\bibitem{Chiba04}
D. Chiba, Y. Sato, T. Kita, F. Matsukura, and H. Ohno, 
cond-mat/0403500.

\bibitem{Yang04}
T. Yang, T. Kimura, and Y. Otani, preprint.

\bibitem{Thi04}
A. Thiaville and Y. Nakatani, private communication.

\bibitem{com0}
 Numerical simulation observed dynamical and chaotic domain pattern 
\cite{Thi04}. 

\bibitem{TF94}
G. Tatara and H. Fukuyama Phys. Rev. Lett. {\bf 72}, 772 (1994); 
J. Phys. Soc. Jpn. {\bf 63}, 2538 (1994). 


 



\bibitem{TK04}
G. Tatara and H. Kohno, Phys. Rev. Lett. {\bf 92}, 086601 (2004).


\bibitem{Rajaraman}
R. Rajaraman, {\it Solitons and Instantons} (North-Holand, Amsterdam, 
1982). 


\bibitem{BL96} H.-B. Braun and D. Loss, Phys. Rev. B{\bf 53}, 3237 (1996). 

\bibitem{TT96}
S. Takagi and G. Tatara, Phys. Rev. B{\bf 54}, 9920 (1996). 

\bibitem{ST00}
J. Shibata and S. Takagi, Phys. Rev. B{\bf 62}, 5719 (2000). 

\bibitem{ST01}
J. Shibata and S. Takagi, Phys. Atom. Nucl. {\bf 64}, 2206 (2001). 

\bibitem{com2}
 In fact, $\Omega^{\rm dw}_{k}$ vanishes when 
$f(x)\equiv (x+1)(x+1+\kappa)-\frac {\kappa^2}{4}x\sin^2 2\phi_0
+2\kappa^2 \sin^2\phi_0\cos2\phi_0=0$, where $x\equiv
(k\lambda)^2(\geq0)$. 
 At $x=0$, $f(0)=1+\kappa+2\kappa^2\sin^2\phi_0\cos2\phi_0$ and is
positive if $\kappa<1$ irrespective of $\phi_0$.
 Noting that $f(x)$ in $x\geq0$ is an increasing function of $x$ for 
$(0\leq) \kappa < 2(1+\sqrt{3})$, we see that $\Omega^{\rm dw}_{k}>0$ 
if $\kappa<1$.


\bibitem{BKL97} H.-B. Braun, J. Kyriakidis and D. Loss, Phys. Rev. B{\bf 56}, 8129 (1997). 

\bibitem{partial}
In this calculation, 
we have performed the integration by parts with respect to $x$ 
in the spin-torque term of Eq.(\ref{effL}) before substituting 
the static solution, $\theta_{\rm s}$ and $\phi_{\rm s}$.  


\end{thebibliography}
\end{document}